\documentclass[runningheads]{llncs}
\usepackage{graphicx}
\usepackage{comment}
\usepackage[backend=biber,style=numeric,sorting=none]{biblatex}
\bibliography{main}
\newcommand{\RQOneandTwo}{How can we model users’ background knowledge, target knowledge and the knowledge gap in knowledge acquisition tasks?}

\newcommand{\RQThree}{To what extent can the incorporation of the knowledge gap into a search system facilitate a more efficient journey for users in knowledge acquisition tasks?}

\newcommand{\RQFour}{How should search systems be evaluated in the context of knowledge acquisition tasks?}

\begin{document}
\title{Relevance Models Based on the Knowledge Gap}
\author{Yasin Ghafourian\inst{1,2}\orcidID{0000-0001-9683-9748}}
\authorrunning{Y. Ghafourian}
\institute{Research Studios Austria FG, Vienna 1090, Austria \and
Vienna University of Technology (TU Wien), Vienna 1040, Austria
\email{yasin.ghafourian@researchstudio.at}}
\maketitle              % typeset the header of the contribution
\begin{abstract}
Search systems are increasingly used for gaining knowledge through accessing relevant resources from a vast volume of content. However, search systems provide only limited support to users in knowledge acquisition contexts. Specifically, they do not fully consider the knowledge gap which we define as the gap existing between what the user knows and what the user intends to learn. The effects of considering the knowledge gap for knowledge acquisition tasks remain largely unexplored in search systems. We propose to model and incorporate the knowledge gap into search algorithms. We plan to explore to what extent the incorporation of the knowledge gap leads to an improvement in the performance of search systems in knowledge acquisition tasks. Furthermore, we aim to investigate and design a metric for the evaluation of the search systems' performance in the context of knowledge acquisition tasks.

\keywords{Information Retrieval  \and Knowledge Delta \and Knowledge Acquisition.}
\end{abstract}
% 

%\paragraph{Preference of Attendance:}Hereby I announce my tendency towards participating in the in-person conference as physical attendance facilitates the interactions and makes the exchange of ideas easier and more motivating. 
\section{Motivation and Problem Statement}

Web search is nowadays considered more to be a source for accessing information resources and exploration, and educational resources are also not an exception to this. Web search has become the most popular medium for education in schools, universities and for professional training \cite{aroyo2006interactive,marchionini2006exploratory}. In addition, web search is being used more often for the aim of gaining new knowledge \cite{gadiraju2018analyzing}. There have been numerous studies done on the information-seeking behaviors of students and academic staff in different parts of the world (some examples are \cite{huda2005information,nkomo2009comparative,igbinoviainformation,gyesi2020information}) which also verify that the majority of individuals and students consider the internet to be the most useful information source for learning.
%As a case in point, the study done by Al-haddabi \cite{huda2005information} surveying 241 students of different educational stages (undergraduate, postgraduate) concludes that the majority of surveyed students ($54.7\%$) use the internet as the most useful information source.

Marchionini \cite{marchionini2006exploratory} categorises the search activities into two broad categories of \emph{look-up} and \emph{exploratory} search activities. Look up search is the activity in which users know in particular what information they want and have a concrete expectation of what would the desired search results be, while in exploratory search users go through multiple iterations of searching the online resources. With regard to using search systems for learning purposes, look-up search is viewed as a type of search that users initiate based on their current knowledge of the subject of interest that leads them to the relevant neighborhood of information. From this neighborhood, users will then start their exploration into the learning resources and might reformulate their query multiple times according to their findings with the aim of reaching more satisfying resources. 

%During a search session initiated to acquire new knowledge on a topic in a given field, users assess the utility of the suggested documents on a search engine's results page (SERP) to figure out which of the documents are worth investigating (consuming) for their purpose of learning about the topic.
Users typically have different levels of background knowledge on the topic in question.
We define a user's background knowledge as the current level of the user's familiarity with the topic. 
Due to different levels of background knowledge, users might perceive the relevance of the documents on a topic differently.
%While one user might find a particular document useful to satisfy their learning goal,  another user might find the same document not so relevant due to their specific background knowledge.

Search engines currently do not take this diversity in the background knowledge levels of the users into account and assume that users' information needs are well represented in their queries \cite{kelly2002user}. This is a clear limitation. The optimal sequence of documents leading to satisfying a user's learning goal depends on the user's specific background knowledge. What if search engines could exploit information about the user's background knowledge to provide the suitable (sequence of) documents helping the user to satisfy their learning goal as fast as possible.  

%The problem is widely important: everyone searches, this is key in solving
As search engines are being used by students and researchers,
%There have been a number of studies that have investigated to what extent the students use the search engines for their studies. As a case in point, the study done by Al-haddabi \cite{huda2005information} surveying 241 students of different educational stages (undergraduate, postgraduate) concludes that the majority of surveyed students ($54.7\%$) use the internet as the most useful information source. Similarly, in another study done by Apuke et al. across three universities \cite{apuke2018university} in Nigeria with 250 participants who were senior year undergraduate students, it turned out that $89.6\%$ of students use the internet for their academic career with $62\%$ of them using it on a daily basis. Considering these studies,
it is crucial for search engines to pursue more developments in this direction so as to be a better fit for learning tasks, especially \emph{knowledge acquisition tasks}. We define \emph{knowledge acquisition tasks} as tasks in which users aim to acquire knowledge with a learning goal as part of a sequence of interactions with an information retrieval (IR) system. Supposing that we have a search system specialized for knowledge acquisition tasks that takes the users' knowledge level into account, it will provide the users with resources that best fit their learning needs according to their knowledge level. This, in turn, saves effort from users in terms of spending longer search times and consuming documents that are returned as relevant but cannot be utilised by the users according to their knowledge level.
%Consequently, this adaptation of results makes such a search system more practical and faster for acquiring knowledge for a greater population of users as these systems will be assisting users throughout the learning process.

% What are we going to do in a nutshell
To incorporate the users' knowledge level into the search, one needs to enable the search system to have a model representing the users' knowledge within the topic of interest (``what the user knows''). Furthermore, one needs to define and represent the knowledge to be acquired (``what the user wants to know''). The goal is then to overcome the gap between these two representations. We call this gap the \emph{knowledge gap}. Having defined the knowledge gap, the objective is to develop a retrieval method that provides users with an order list (a path) of resources,
%in a pedagogical order, 
that helps them to overcome the knowledge gap. A suitable order will be one where more complex resources requiring more background knowledge will be preceded by resources that are more easily approachable based on the user's background knowledge. In this research, we will investigate the means to measure the \emph{knowledge gap} and understand how one applies it within a search system designed for acquiring knowledge so that the users can effectively overcome the knowledge gap. In the rest of this paper, we will first discuss the research questions that we seek to answer throughout this research. Later and in the section that follows, we will provide a brief overview of the research surrounding the concepts of the \emph{knowledge gap} and the \emph{knowledge delta}. Finally, we will explain our planned methodology to approach the research questions.

\section{Research questions}
Our motivation is to investigate on how to improve the retrieval effectiveness of the search systems for knowledge acquisition by incorporating the knowledge gap into the ranking mechanisms. We propose our main research question as follows:

\textbf{High level research question:}
How can the search system help users to effectively overcome their knowledge gap in a knowledge acquisition task?

This high-level research question is comprised of three fine-grained research questions to be investigated:

\begin{itemize}
    \item RQ1: \RQOneandTwo
    %\item RQ2: \RQTwo
    \item RQ2: \RQThree 
    \item RQ3: \RQFour
\end{itemize}

\section{Background and related work}\label{sec:litreview}
In this section we provide the result of our literature review done in an attempt to capture the viewpoint of the papers that discuss the concepts of \emph{knowledge gap} and \emph{knowledge delta} and how they approach and incorporate these concepts. 

\subsection{Knowledge Gap}

What we defined earlier, i.e the gap between the users' knowledge level and the level of knowledge in the field of interest for learning has been discussed in the literature under the title of \emph{knowledge gap}. Knowledge gap is one of the causing factors of information need \cite{cosijn2000dimensions}. Another factor that causes information need is referred to as \emph{Anomaly in the state of knowledge} by Belkin et al. \cite{belkin1982ask} which is the phenomenon in the state of knowledge that causes the information need. In one of the early studies on information use, Dervin and Nilan \cite{Dervin1986information} used the phrase \emph{knowledge gap} to refer to a situation where a person's cognitive state has recognized an incompleteness in its currently possessed information. This incompleteness happens as a result of interaction with information sources or through thinking processes, and thereby later that incompleteness will turn into an information need. Additionally, Thellefsen et al. \cite{thellefsen2014pragmatic} use \emph{knowledge gap} and \emph{information need} interchangeably in a discussion where some of the definitions of \emph{knowledge gap} are covered to argue the intricacy of the concept of information need and the importance of incorporating users' information need while developing a knowledge organization system. In a research done by Yu et al. \cite{yu2021topic} the \emph{knowledge gap} also has a similar explanation, however, the research is more focused on knowledge predictive models for the knowledge state of users which are being calibrated through questionnaires. Considering the definition of \emph{knowledge gap}, there are studies that use the same definition and provide search solutions that are adaptive to the \emph{knowledge gap} \cite{stojanovic2005role,soldaini2019knowledge,zhang2020users}. In addition, there are similar studies that seek to model the knowledge gap by modeling the knowledge of the user and compare it against an existing knowledge level for the topic of interest \cite{aroyo2006interactive,xiong2020automated,thaker2020recommending,lindstaedt2009getting,syed2020models,zhao2016paper}. Among these works, the work done by
Zhao et al. \cite{zhao2016paper} explores the knowledge paths between users' already acquired knowledge and the target knowledge that the user aims to obtain. Considering this knowledge path and in the context of recommender systems, the authors' method recommends a number of papers to the users so that their learning goals are achieved in the best satisfying way. Similarly, as our goal will be to assist users in knowledge acquisition tasks, we will take into consideration the target knowledge that users aim to obtain.  Having modeled the knowledge from users' background knowledge and the target knowledge level, our goal is to estimate the knowledge gap between these levels. Thereafter,
we will use the modeled knowledge gap to improve the retrieval effectiveness of the search systems for knowledge acquisition.
%we will investigate on how to improve the retrieval effectiveness of the search systems for knowledge acquisition by incorporating the knowledge gap in the ranking mechanisms.

\subsection{Knowledge Delta}
\emph{Knowledge delta} is another concept that is semantically closely related to knowledge gap.
One definition of \emph{knowledge delta} in the literature is the amount of knowledge change in a user's knowledge level which can be measured through questionnaires. As a case in point, we can refer to the work of Grunewald et al. \cite{grunewald2013designing}, where the expertise gain is calculated through asking users of a "Massive Open Online Courses" system about their knowledge level in a field before and after taking an online course and denoting it as \emph{knowledge delta}. Similar studies have also been done to measure the knowledge change of the users \cite{grunewald2013designing,daghio2006assessment,daghio2006evaluation,schaumberg2015variation,de2020introducing}.

In all the aforementioned works, \emph{knowledge delta} is used as a concept that demonstrates the change in users knowledge level. This concept is also associated with the name of \emph{knowledge gain} in other works such as \cite{vakkari2019modeling}. 
%From the perspective of this proposal, however, \emph{knowledge delta} and \emph{knowledge gain} are different. \emph{Knowledge delta} is the actual knowledge gap that exists between what a user knows and what the user wants to know and the change in the user's knowledge level which is denoted as \emph{knowledge gain}, doesn't necessarily fulfil this knowledge gap. In the next subsection, we look closer for conducted research articles that address the gap between a user's knowledge level and the knowledge surrounding the user's topic of interest.

\section{Research Methodology}
The methodology begins with the investigation to find an answer to the first research question all the way to the third research question. The steps of the methodology are three-fold. Firstly, we will build a model for a user's knowledge and use it to build a model for the user's knowledge gap. Secondly, we will ask users to participate in knowledge acquisition tasks to gain knowledge on learning goals that will be defined for them. During this experiment, we will utilize the modeled knowledge gap for each user during the user's search session in order to provide the user with better results for learning. during this step, we will investigate the extent of improvement that incorporating the knowledge gap will bring about in a learning session. Thirdly, we will design a function whose output is a measure that will score the user's learning progress throughout the session.
%This function's goal will be to introduce a session-based evaluation metric that gauges the quality of learning in knowledge acquisition tasks.
%An overview of the methodology is demonstrated in figure \ref{Methodology_Unified}.

\begin{center}
\textbf{\textit{RQ1: \RQOneandTwo}}
\end{center}

%The aim of the first research question is to investigate how we can model a user's knowledge and knowledge gap. To achieve this end,
To answer this research question,
we need to design learning tasks for the users and we need to estimate the users' domain knowledge.
Each learning task will set gaining knowledge on a sub-topic within a topic as a goal for the users.
%Each learning task will involve gaining knowledge on a sub-topic within a topic. As a result, learning about the sub-topic will be the learning goal for the users.
This goal will establish a desired level of knowledge as the target knowledge that a user wants to acquire. We will choose a fixed number of sub-topics and assign each user to a sub-topic for the learning task. 

As it was mentioned under section \ref{sec:litreview}, several studies have modeled a user's knowledge in a variety of tasks and in different ways \cite{aroyo2006interactive,zhang2020users,zhao2016paper,yu2018predicting,zhang2015predicting}. Extending the work done by Zhang et al \cite{zhang2015predicting}, we aim to represent the user's knowledge of a topic with a set of concepts within that topic . 

We will assess the user's knowledge before and after the learning session using knowledge tests.
%and knowledge tests.
We will design the knowledge tests in such a way that the user's knowledge of each concept can be attributed to a level of learning according to Bloom's taxonomy of educational objectives \cite{bloom1956taxonomy,anderson2001taxonomy}.
%In addition to the questionnaires, we will also design knowledge tests. A knowledge test, which is a set of questions investigating the knowledge in concepts of a domain, will also be designed in accordance to Bloom's taxonomy. As a result, knowledge tests will provide an estimation of the user's domain knowledge in the context of a certain learning level of Bloom's taxonomy.
%A user will be asked to take a knowledge test.
%In this questionnaire, the user will be guided to choose an appropriate level of understanding for their knowledge about each concept.
%The user will also be asked to take a pre-task knowledge test. 

Having completed the knowledge test,
%and the knowledge test phases
each user's knowledge will be represented with its set of constituent concepts. There are a variety of options to represent the user's knowledge. One such way is a one-dimensional vector of concepts and the user's understanding of the concepts as weights. %One option could be to represent them as a vector of weights with weights corresponding to Blooms' learning levels.
%Another option could be to build a concept tree where the concepts are the the nodes of the tree with each node having its weight similar to the concept vector representation. In this tree representation, the branches of the tree will be built as a prerequisite chain of concepts \cite{Wang2016,Gordon2018,Manrique2019} according to the concepts' pedagogical role. A prerequisite chain of concepts demonstrates that, within a topic, learning advanced concepts requires already having learnt a number of prerequisite, easier concepts.
%Another option to represent the user's knowledge could be to use a two dimensional space. In this space, concepts form one dimension of the space and the user's understanding of those concepts will be the other dimension.
%The model adopted to represent the user's knowledge,
The same model will also be used to represent the target knowledge. The knowledge gap will then be computed between these two models.

Having modeled the knowledge gap, the methodology continues the investigation by moving to the second research question. 

\begin{center}
\textbf{\textit{RQ2: \RQThree}}
\end{center}

%Our assumption is that by being aware of the knowledge gap, we will be able to create a \emph{personalized understandability metric}. This metric will be used to give a personalized understandability score to learning resources. This score will be used along with relevance scores to present the best learning resources to the user. 

After the pre-task knowledge assessment, users will have access to a search interface connected to a search engine (here after collectively referred to as the search system) to carry out their learning tasks.
%which could be based on either Elasticsearch\footnote{https://www.elastic.co/} or Pyserini\footnote{https://github.com/castorini/pyserini}. The search engine will have indexed the datasets of learning resources for the topics of learning tasks. The purposes of having a search interface as an intermediary between the user and the search engine are two-fold. Firstly, 
The interface will allow for the recording of the users' interactions which will later be used as features to evaluate the quality of the learning sessions.
%Secondly, the interface will allow for more flexibility in the manner of presenting the suitable learning resources to the user.
%In this experiment, We call the search engine and the interface together as the search system.

Each user will participate in at least two learning sessions during this experiment. In one learning session, the search system will not take the knowledge gap into account and original retrieved results by the system will be presented to the user. In the other learning session, the search system will take the knowledge gap into account and adapt the results before presenting them to the user. There are several ways to adapt the search results to the knowledge gap. One such way will be to re-rank the results of the user's each query submission based on a personalized understandability score of those results.
%Another way will be to provide an option to the user in the search interface to be able to modify their level of knowledge known by the system. We will show the 'concept' -'level of understanding' pairs to the user where the user can declare a new level of understanding for each concept. Each time users make changes to the knowledge stored in the system as sign of learning, new resources will be shown to the user.
For each of the learning sessions, we will design a different learning task. So as to ensure that learning about one topic doesn't affect the user's knowledge about other topics.
%As a result, the learning outcome of the learning sessions for each user will be independent. 
After the learning session, the users will be asked to take a knowledge test again (Post-task knowledge tests). Comparing the result of the pre-task knowledge test with the post-task test for each user, we will have a score for the progress of the user's knowledge. On the other hand, for each user, we will compare the knowledge gained in the learning sessions to observe the effect of incorporating the knowledge gap on the quality of the learning session. We will evaluate this quality based on the interaction features recorded during the session. %In the study done by Yu et al. \cite{yu2018predicting} to predict the knowledge gain of the users, authors identified a set of 70 interaction features. The authors further showed that a number of features, such as 'number of unique pages visited during the session' or 'maximum time spent on page', have more correlation with the amount of knowledge gained. We will also use these features selected by Yu et al. \cite{yu2018predicting} to compare between the experiments. 
%By comparing the self-announced knowledge through questionnaires with the results of the knowledge tests for each user, we expect to see the Dunning-Kruger effect \cite{kruger1999unskilled} in the user's knowledge level. This effect is a phenomenon in which people with limited knowledge or competence in a given intellectual or social domain greatly overestimate their own knowledge or competence in that domain relative to objective criteria.
%Having analyzed the extent of improvement that incorporation of the knowledge gap has in the knowledge acquisition tasks, the second research question concludes.
%The methodology will further proceed towards investigation for the third research question.
\begin{center}
\textbf{\textit{RQ3: \RQFour}}
\end{center}

In this research question, we will investigate how to define evaluation performance metrics for search systems in knowledge acquisition tasks. Previously in \cite{ghafourian2021information}, we have explained why current retrieval metrics are insufficient in this context as they treat each query during the session independently and don't consider the users' knowledge factor in the evaluations. Correspondingly, we proposed three directions forward as well as their advantages and shortcomings: 1) online evaluation approach, 2) prerequisite-labeled relevance judgements approach, and 3) session-based evaluation approach. It's essential to mention at this point that what has been defined in this research proposal as \emph{knowledge gap} was defined as \emph{knowledge delta} in \cite{ghafourian2021information}. However, in order to maintain a greater consistency with the literature in this area considering the subtle distinction between the \emph{knowledge delta} and the \emph{knowledge gap}, we have adopted the term \emph{knowledge gap} in this research proposal. %The evaluation approaches proposed in \cite{ghafourian2021information} are briefly described below. 

%Firstly, an \textbf{online evaluation approach} which suggests comparing a system that incorporates the knowledge gap with a system that doesn't in terms of online evaluation metrics such as Click through rate (CTR). 

%Secondly, a \textbf{Prerequisite-labeled relevance judgements approach} which suggests enabling the offline evaluation of systems empowered by considering the knowledge gap through creating a relevance judgment file that incorporates the background knowledge assumed for each relevance judgment record to hold true. 

%Thirdly, a \textbf{Session-based evaluation approach} in which the goal is to look for the shortest sequence of queries and accessed documents that will lead the user to the state in which the user overcomes the knowledge gap and becomes familiar with the target topic.

Our objective will be to extend this previous work and formalise the \emph{session-based evaluation approach}. The goal is to design a session-based evaluation function that gauges the quality of learning sessions in knowledge acquisition tasks.

Up until this point in the methodology, we will have collected data on knowledge that the users have gained and information about learning sessions. %In other words, we will have gained information about learning sessions during which users have had greater knowledge gains according to their knowledge tests.
In addition we will have recorded the users' interactions with the search system. As a result, a function that will serve as a new performance metric will be designed. This performance metric will use interaction features, such as time, number of queries used, etc.,  as cost indicators. It will combine these cost indicators such that the function's value aligns well with the experiences from the qualitative data (Knowledge tests). As a result, for the evaluation of future learning sessions, it will suffice to only have access to cost indicators and to use the designed function to evaluate the learning session.
%Thereby, there will be no need for qualitative data.

There two main research challenges for the implementation of the discussed methodology: 
\begin{enumerate}
    \item 
    The challenge that exists for the implementation for research question two is maintaining a balance between guiding a user to resources that are better suited for them and are adapted to their level of knowledge by the system, and just responding to the queries the user submitted. %If the results presented to the user diverge so much from the target knowledge of the users as a consequence of the additional re-ranking step, and the users start to feel that they don't have the power to control the response of the search system, the experiments will have invalid outcomes.
    \item In defining the learning goal for the experiments, the level of understanding of the user in the topic is a variable and the desired change in the level of understanding for the sake of experiments should be fixed. (e.g. Does it suffice if users are just familiarized with a concept, or should they be able to use it or should they be able to implement it?)
\end{enumerate}

\section*{Acknowledgements}
  This project has received funding from the European Union's Horizon 2020 research and innovation programme under the Marie Skłodowska-Curie grant agreement No 860721.

\printbibliography
\end{document}